\documentclass[twocolumn,epsfig,pre]{revtex4}

\usepackage{graphicx}
\usepackage{graphics}
\usepackage{amsmath}
\usepackage{amssymb}
\usepackage{hyperref}
\usepackage{latexsym}

\def\be{\begin{equation}}
\def\ee{\end{equation}}
\def\bea{\begin{eqnarray}}
\def\eea{\end{eqnarray}}

\draft

\begin{document}
\title{Fluctuation induced chiral symmetry breaking in autocatalytic reaction-diffusion systems}
\author{V. S. Gayathri$^{1}$ and Madan Rao$^{1,2}$}
\affiliation{
$^1$Raman Research Institute, C.V. Raman Avenue,
Sadashivanagar, Bangalore 560080,
India\\
$^2$National Centre for Biological Sciences-TIFR, UAS-GKVK Campus, Bellary Road, Bangalore 560065, India}
\begin{abstract}
We show how spatiotemporal fluctuations can induce {\it spontaneous symmetry breaking} in systems which are perfectly symmetric in the absence of fluctuations.
We illustrate this in the context of the autocatalytic production of chiral enantiomers from achiral reactants
in reaction-diffusion systems. The mean field steady state is chiral symmetric; spatiotemporal fluctuations induce a novel (molecular) chiral ordering and sharp phase transitions including reentrance. 
We discuss its implications in the context of the emergence of molecular homochirality.
\end{abstract}


\maketitle

Fluctuations are known to change the nature of phase transitions or even destroy long-range order, and so it comes as a surprise that
fluctuations may induce {\it symmetry breaking transitions} in systems where such symmetry breaking would not have occured in its absence \cite{note1}.  In this Letter we
study the autocatalytic production of chiral enantiomers from achiral reactants
in reaction-diffusion systems whose mean field steady state is chiral symmetric; spatiotemporal fluctuations however induce a novel chiral ordering and sharp phase transitions.

While arriving at these results we comment on 
the conventional understanding of homochiral amplification in autocatalytic systems. F.C. Frank \cite{frank} proposed a simple 
chemical kinetics model for the amplified production of a chiral enantiomer from a racemic mixture seeded with an initial bias, based on {\it autocatalysis} of the favoured enantiomer and {\it antagonism} of the disfavoured. 
Following recent successful realisations of this scheme \cite{soai}, it is widely believed that both an initial chiral seed and antagonism are necessary conditions for the amplification of homochirality in autocatalytic systems \cite{blackmond}. Contrary to this, we show
that {\it complete chirality discrimination and exponential amplification} can occur, (i) without an initial chiral seed and (ii) without specific antagonism, simply via stochasticity inherent in autocatalytic chemical reactions.

Consider a system of achiral molecules $A$ and $B$ which diffuse and react on collision to form chiral enantiomers $R$ and $R^{*}$,
\begin{eqnarray}
\label{auto}
  \begin{array}{ccc}  \, & {\scriptstyle {\,_{k_1(R)}}}
& \, \\
A + B & {\rightleftharpoons} & R \\
\, &  {\scriptstyle{\,^{k_2}}} & \, \\
 \, & {\scriptstyle {\,_{k_1(R^{*})}}}
& \, \\
A + B & {\rightleftharpoons} & R^{*} \\
\, &  {\scriptstyle{\,^{k_2}}} & \, 
\end{array}
 \end{eqnarray}
The forward reactions are autocatalytic with rates $k_1(R)$ and $k_1(R^*)$ -- existing enantiomers favour their own production -- while the backward reactions with identical rates $k_2$ are non-autocatalytic. 
A simple linear model of autocatalysis is $k_1(R) = k_1 P(N_R)$, where $P(N_R)$ is the probability of encountering 
$R$ in the forward reaction. In our symmetric autocatalytic model, we take the constants $k_1$ to be the same for the two reactions. The reverse reaction, on the other hand,  has no reason to be autocatalytic; even so our qualitative conclusions remain the same were the reverse reactions autocatalytic too. Note that the molecules interact only through the autocatalytic chemical reactions, which occur on collision. 

First define the mean-field model for this autocatalytic reaction-diffusion system, valid when the number of molecules per unit volume is large so as to have a well-defined concentration field which varies smoothly and deterministically with space and time. The mean-field condition implies that local fluctuations in the concentration are much smaller than the mean, $\sqrt{\bar{n^2} - {\bar n}^2} \ll {\bar n}$. Under these conditions we may replace the probability by the concentration of the reactant at time $t$, expressed in
number of moles per m$^3$,
and describe the mean field chemical kinetics of (\ref{auto}) by reaction-diffusion equations such as,
\begin{equation}
\label{reAB}
{\partial_t n_A} =   - k_1 n_A n_B \left(n_R + n_{R^*} \right) + k_2 \left(n_R + n_{R^*}\right)
+ D_A \nabla^2 n_A
\end{equation}
and
\begin{equation}
\label{reR}
{\partial_t n_R}     =     k_1 n_A n_B n_R - k_2 n_R 
+ D_R {\nabla^2 n_R}
\end{equation}
(similar for $R^*$). The rates of reaction, $k_1$ and $k_2$ have units of m$^6$s$^{-1}$ per mole$^2$ and s$^{-1}$ respectively. 

These equations can be solved exactly in the well-stirred limit (drop spatial derivatives) to obtain
$d n_R/d n_{R^*} = n_R/n_{R^*}$. This implies that 
with symmetric initial conditions, $n_R(0) = n_{R^*}(0)$, there can be 
no chiral symmetry breaking, i.e., $n_{R}(t) = n_{R^*}(t)$ at all times.
A fixed point analysis of (\ref{reAB}) and (\ref{reR}) shows that there is a line of neutral stability with $n_A=n_B = \sqrt{k_2/k_1}$ --- there exist marginal fixed points corresponding to both chiral symmetric and asymmetric states. This implies that an arbitrary perturbation from the chiral symmetric fixed point generated by the slightest stereo-preference, will remain as is (i.e., unamplified) and will not give rise to a {\it global chiral symmetry breaking} \cite{noteamp}.

The same analysis may be applied to the reaction-diffusion equations (\ref{reAB}),\,(\ref{reR}); if the local 
concentration field is large enough, $n_A \geq \sqrt{k_2/k_1}$, there is no symmetry breaking. We next study the effect of including an {\it additive} or a {\it multiplicative} noise in  (\ref{reAB}) and \,(\ref{reR}). Such sources of noise arise in an {\it open} chemical system through spatiotemporal fluctuations in the number or temperature or indeed any other `fast' variable that couples to the chemical rates. The 
additive (multiplicative) noise is
distributed independently and uniformly with zero (nonzero) mean and variance chosen such that the concentrations are non-negative at all times. We have numerically checked that such an implementation of stochasticity does not lead to chiral symmetry breaking starting from {\it symmetric} initial conditions. Indeed in multiplicative noise models, it is easy to see that the zero relative enantiomeric concentration $\phi \equiv  n_R - n_{R^*} = 0$ is an {\it absorbing} state \cite{bykov}.

An important feature of mean-field chemical kinetics
is that in every interval $\Delta t$,
all reactions take place simultaneously. Thus the stochasticity
considered so far is incapable of generating global chiral symmetry breaking from symmetric initial conditions.

Chemical reactions however have another source of inherent stochasticity,
since chemical changes occur during molecular 
encounters which are fundamentally discrete stochastic processes; the probability of a reaction occuring is related to the probability of the reactants encountering each other in a time interval $\Delta t$ derivable from kinetic theory. 
This suggests that the appropriate stochastic description is to treat the reaction-diffusion kinetics as a markov process described by a master equation for the probability distribution of the number of molecules of a given species within a 
`local' coarse-grained volume. This treatment should reduce to the mean-field model
when the number of molecules per unit volume is large enough to have a well-defined concentration field varying smoothly and deterministically with space-time.

We first construct a local coarse-grained volume by noting that at scales smaller than $\xi \sim \sqrt{D t_{r}}$, where
$D$ is the typical diffusion coefficient and $t_{r} \sim (k_1 {\bar n}^2)^{-1}$ is the typical reaction time, the chemical system may be considered well-stirred. To incorporate spatial fluctuations
over scales larger than $\xi$, we divide the chemical pool into $M$ blocks, each of size $\xi$ situated on a cubic (square) lattice. The contents in each box is assumed well stirred, however each box is connected to its $z$ nearest neighbours allowing the reactants to diffuse into neighbouring boxes. The evolution of the probability distribution of the number of molecules of a given species within each box, is specified by writing down transition probabilities obeying detailed balance, which involve both reaction processes in each box and diffusion processes connecting nearest neighbour boxes.
An efficient stochastic algorithm for solving the reaction master equation is the
Gillespie algorithm \cite{gillespie}, which takes as inputs,
the initial numbers of each species and the chemical rate constants $k_1=c_1 \xi^{2d}$ and $k_2$. The number of molecules of each species is  updated whenever a reaction (following a collision) occurs; this is a stochastic event and the probability of its occurence is computed from kinetic theory.

Diffusion can be modeled by transferring a molecule of a given species from box ${i}$ to a neighbouring box  ${j}$ with a rate given by $\lambda = D/\xi^2$, where $D$ is the diffusion coefficient of the given species \cite{marion}.
Thus if $N_R^{i} (t)$ is the instantaneous number of $R$ molecules in block $i$, then 
we may augment the master equation by a loss term 
\begin{equation}
P\left[N^{(i)}_{R}(t+\Delta t) = N^{(i)}_{R}(t)-1\right] = \lambda 
N^{(i)}_R(t)\,  \Delta t  \nonumber
\end{equation}
and a gain term 
\begin{equation}
P\left[N^{(i)}_{R}(t+\Delta t) = N^{(i)}_{R}(t)+1\right]  =  \frac{\lambda}{z} \sum_j N^{(j)}_R(t) \, \Delta t\, .  \nonumber
\end{equation}
We have explicitly checked that this correctly describes the diffusive dynamics of $R$; in the continuum limit this recovers the particle current $J \propto \nabla n$. 

In each box we fix a different random number seed and initial data set $\{N^{(i)}_A = N^{(i)}_B, N^{(i)}_R = N^{(i)}_{R^{*}}\}$. 
The time evolution of physical observables are computed from the stochastic master equation; in all displayed
graphs we have chosen $c_1=10^{-3}$s$^{-1}$ and $k_2=10^{-1}$s$^{-1}$ and have checked that our qualitative results remain unaltered when we vary the rates $c_1 = 10^{-3} - 10^{-5}$ and $k_2 = 10^{-1} - 10^{-2}$ (changing these rates merely changes the time and space scales).
We have made time $t$ dimensionless by multiplying with $k_2$. 
Averages denoted by $\langle \ldots \rangle$ are over 10 realisations of random number seeds, further the temporal behaviour of various observables are smoothened by binning over a time interval of $t=5$.

First the diffusionless limit, $\lambda =0$ : the boxes are uncoupled and independent of each other.
We follow the reaction dynamics in a typical box; the initial data and the chemical rates are perfectly chiral symmetric, further recall that the mean field steady state is also chiral symmetric. Despite this, in the stochastic evolution the populations of $R$ and $R^{*}$ stay roughly the same for a while before exhibiting a definite chiral symmetry breaking beyond a time scale $\tau$ (Fig.\,1). Subsequently, the population of one of the enantiomers, say $R^{*}$, goes to zero, while the other rapidly takes over the entire population, $\langle N_R \rangle \propto 1-\exp(- \beta t)$, with a rate that is an increasing function of $k_1$. 
The average number of $A$ (and $B$) molecules decrease with time to an asymptote which is a weak function of initial conditions. 

Chiral symmetry breaking is best expressed by the behaviour of
the relative enantiomeric excess $\Phi = \left\langle{\frac{N_R - N_{R^{*}}}{N_R + N_{R^{*}}}}\right\rangle$ which shows a clear {\it bifurcation} before saturating to $\pm 1$ (Fig.\,1(inset)).
We compute $P(\tau)$, the distribution of the time scale  beyond which chiral symmetry breaking occurs (Fig.\,2) : $P(\tau) \sim \tau \exp (- \tau/\tau_0)$ with a mean $\langle \tau \rangle$ that scales linearly with $N$, for large $N$.
Since chiral symmetry is broken in every box, the saturation value of the spatially averaged $\overline{\Phi}(t=\infty) = M^{-1} \sum_i \Phi_i (t=\infty) \equiv \overline{\Phi}$ is zero.

This dynamical bifurcation may be readily understood in a simplified reaction scheme, 
$A \rightleftharpoons R$, $A \rightleftharpoons R^{*}$, with the forward autocatalytic reaction rate
given by $k_1 n_{R}/(n_{A}+n_{R}+n_{R^*})$ and $k_1 n_{R^*}/(n_{A}+n_{R}+n_{R^*})$ respectively,  where it is  easy to show the following \cite{future} :
(i) the mean field equations do not show chiral symmetry breaking, (ii) the master equation shows chiral symmetry breaking and (iii) the master equation goes over to the mean field equations in the limit of large number of reactants.

Turning on a small $\lambda$ couples the boxes to one another.
The steady state of this coupled map lattice depends on two time scales --- the typical chirality breaking time $\tau$ and the
diffusion time $\tau_d (\lambda)$. Clearly when $\tau_d \gg \tau$, i.e., small $\lambda$, each asymptotic $\Phi_i $ is still $\pm 1$ with equal probability leading to the order parameters, $\overline{\Phi} = 0,\, \overline{\Phi^2} = 1$, a locally chiral but globally racemic state. 

At intermediate values of $\lambda$, when $\tau \sim \tau_d$, one might expect a global chiral
symmetry broken state; that this is so is seen from Fig.\,3 --- the magnitude of the order parameter $\vert \overline{\Phi} \vert$
shows a distinct ``jump'' at $\lambda^{*}$, which gets sharper as $M$ is increased.
We have checked that this symmetry broken state persists by simply running the simulation for longer times. Once the system has locked into this symmetry broken state, fluctuations arising from diffusive transport can no longer restore the symmetry.
Note as in the dynamics towards any symmetry broken state (e.g., a quench from the paramagnet to the Ising ferromagnetic state), an average over different random seeds averages over the two degenerate symmetry broken configurations. The evolution of the order parameter with time shows a typical nucleation profile (Fig.\,4), with an initial lag time and
a subequent exponential rise after a nucleation time $\tau_n$ which is independent of system size (indicating a finite correlation length at the transition) and increases with increasing $\lambda$. 
The stability of this symmetry broken phase can be tested by creating a `droplet' of size $L$ containing the opposite
phase; we find that the mean time for droplet `evaporation'  $\langle \tau_{e} \rangle$ scales as $L^2$ (Fig.\,4).

To understand this novel symmetry breaking transition, we coarse-grain our lattice of boxes into cells of
size $\xi^{\prime}$, the coarse grained $\Phi^{\prime} = m^{-1} \sum_{i=1}^{m} \Phi_i$ where $m=\xi^{\prime}/\xi$. At steady state, $\Phi^{\prime}$ takes values $0, \pm 1$. In terms of these coarse-grained variables, the equations for $\Phi^{\prime}$ is given by
\begin{equation}
\label{coarse}
\frac{\partial \Phi^{\prime}}{\partial t} = D {\nabla^{\prime}}^2 \Phi^{\prime} - V'(\Phi^{\prime}) + \eta^{\prime} (x^{\prime},t)
\end{equation}
where $V' \equiv \partial V/ \partial \Phi^{\prime}$ is derivable from a `fluctuation-induced potential' of the form $V = r {\Phi^{\prime}}^2 - u {\Phi^{\prime}}^4 + v {\Phi^{\prime}}^6$, with positive coefficients. Discretising the laplacian converts (\ref{coarse}) into a coupled map lattice; if fluctuations induce one box 
to reach a chiral state, say $R$, then diffusion of $R$ from this box to its racemic neighbour will explicitly bias the neighbouring box to $R$.  Thus the diffusion $D$ is the ordering influence, leading to a 
discontinuous symmetry breaking when large enough. 

At the other extreme of large diffusivity, $\lambda \to \infty$, the time scales $\tau \gg \tau_d$, i.e., the molecules make rapid excursions to other boxes. Thus in the limit of large $M$, the system behaves as a single box with a large number $N=\sum_i N_i$ of molecules. This is the mean-field limit where all the reactions happen simultaneously, the fluctuations leading to chiral symmetry breaking are absent. Thus, we expect the steady state to be $\overline{\Phi} = 0,\, \overline{\Phi^2} = 0$, a local (and global) racemic state.
We believe that the transition from this chiral symmetry broken phase to the racemic phase at larger $\lambda = \lambda_c$ is sharp and discontinuous, as can be seen in the simpler reaction scheme, 
$A \rightleftharpoons R$, $A \rightleftharpoons R^{*}$ \cite{future}. Thus we claim that the steady state diagram as a function of $\lambda$, shows a chiral symmetry broken phase at intermediate $\lambda$ and a {\it reentrant} racemic phase at low $\lambda$, separated by a sharp discontinuous phase boundary. 

The underlying reason for the chiral symmetry breaking in this autocatalytic reaction comes from the stochasticity in molecular encounters --- fluctuations resulting in a run of successive $R$-reactions, will be followed with higher probability by more $R$-reactions (autocatalysis). 
Further, because of the autocatalytic nature of the reactions, 
once a symmetry broken chiral state has been chosen, there are {\it no finite-size fluctuations} which take the system to the other degenerate state. Note that the symmetry broken steady state is {\it not} the equilibrium state, 
which can be obtained from the stationary solutions of the chemical master equation. The equilibrium distribution 
is clearly chiral symmetric, depends only on the ratio
of the reaction rates $k_1/k_2$ and is independent of the diffusion coefficients. This fluctuation induced symmetry breaking transition occurs in finite size systems and is independent of spatial dimensionality.

To summarize, we have identified a new role for fluctuations
in generating a symmetry broken state in systems which are perfectly symmetric in the absence of fluctuations. We illustrate this phenomenon in the context of autocatalytic reaction-diffusion systems : spatiotemporal
fluctuations drive the chemical system to a chiral symmetry broken steady state. 

We end with some observations  ---
(i) The symmetry breaking arising from stochastic switching of the degenerate autocatalytic reactions is a chemical realisation of Parrondo's games \cite{parrondo}.
(ii) Since the diffusion coefficient is a function of temperature, our phase diagram allows for an interesting interpretation in the context
of emergence of homochirality in the primordial chemical pool. At high temperatures, when $\lambda$ is high, the pool is in a local
racemic state; as the primordial pool cools down, so that $\lambda < \lambda_c$, 
global enantiomeric selection occurs. Once selected, homochirality stays on, even when the temperatures reduce further. 
Note contrary to popular belief \cite{blackmond}, antagonism is not a necessary condition for chiral symmetry breaking.
(iii) Is there an analogue of a `symmetry breaking field' by which we may robustly bias the selection of a specific enantiomer ? An obvious way is to explicitly couple the chemical reactions to an external field such as a substrate or circularly polarised radiation which would provide a stereospecific bias. Another possibility is to drive the chemical soup away from thermodynamic equilibrium (e.g., mechanical stirring \cite{kondepudi,metcalfe}), providing a {\it chemical ratchet} capable of rectifying dynamical fluctuations to bias  `motion' in chemical space. 
(iv) The autocatalytic reactions (\ref{auto}) share features in common with the molecular quasi-species model \cite{eigen}, namely, replication ($R \rightarrow 2R$) and mutation ($R \leftrightarrow R^*$). Fluctuation induced selection could be a good candidate for neutral mechanisms of selection in the absence of fitness pressures \cite{kimura}. 

We thank S. Ramaswamy, N. Kumar and especially A. Dhar, C. Dasgupta and S. Sengupta for critical inputs and very useful suggestions which significantly improved our understanding. We are grateful to D. Kondepudi for correspondence and for sending his papers on the subject. MR acknowledges a Swarnajayanthi grant from DST, India.

\newpage

\begin{widetext}

\begin{figure}
\begin{center}
\includegraphics[width=4in]{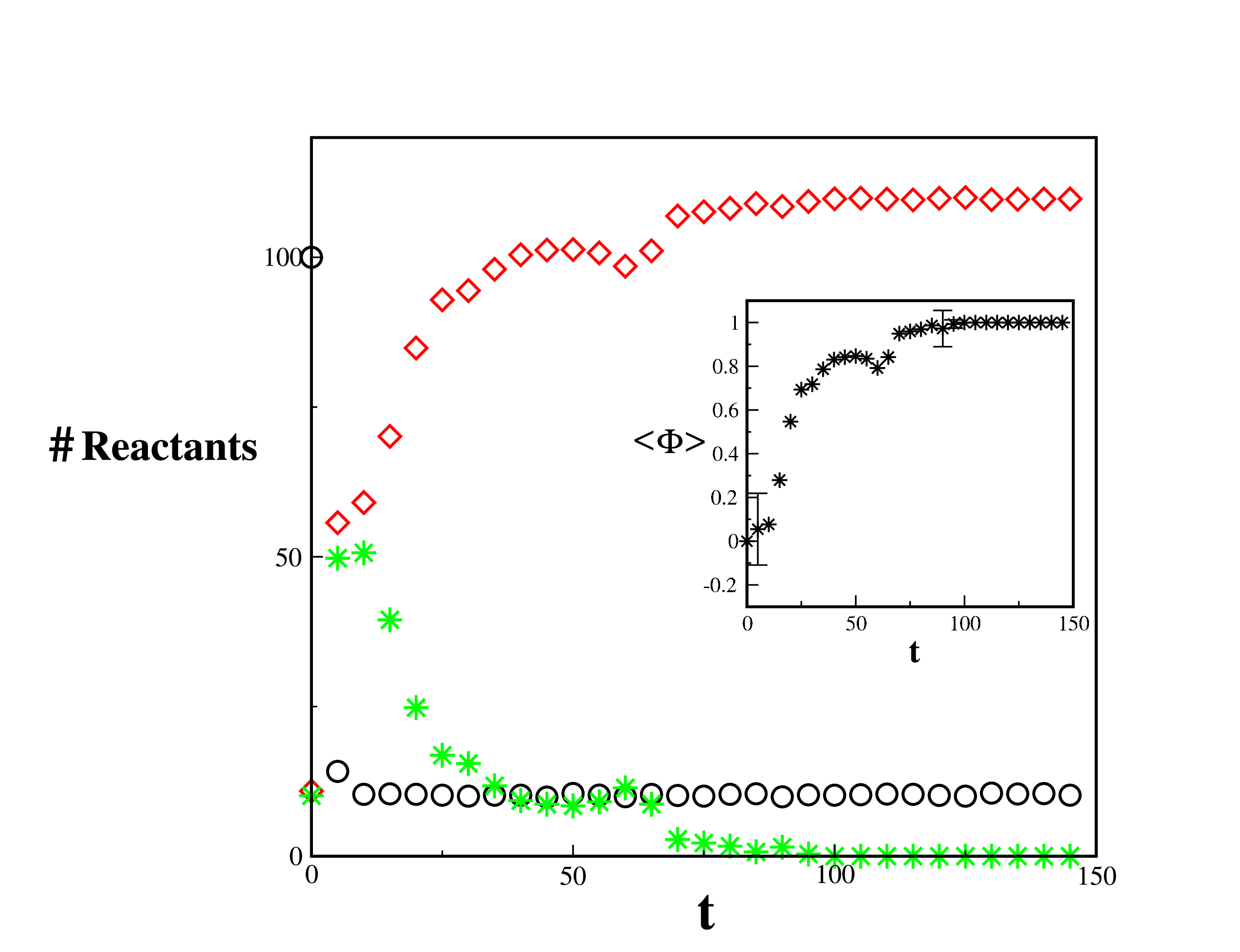}
\caption{Number of reactants versus time for a given set of initial data : while A and B($\circ$)  saturate to a constant population, there emerges a definite enantiomeric selection between R($\diamond$) and R$^*$($\square$). Inset shows the time evolution of the relative enantiomeric excess $\langle\Phi\rangle$, with error bars from different noise realisations.}
\label{reactants}
\end{center}
\end{figure}

\begin{figure}
\begin{center}
\includegraphics[width=4in]{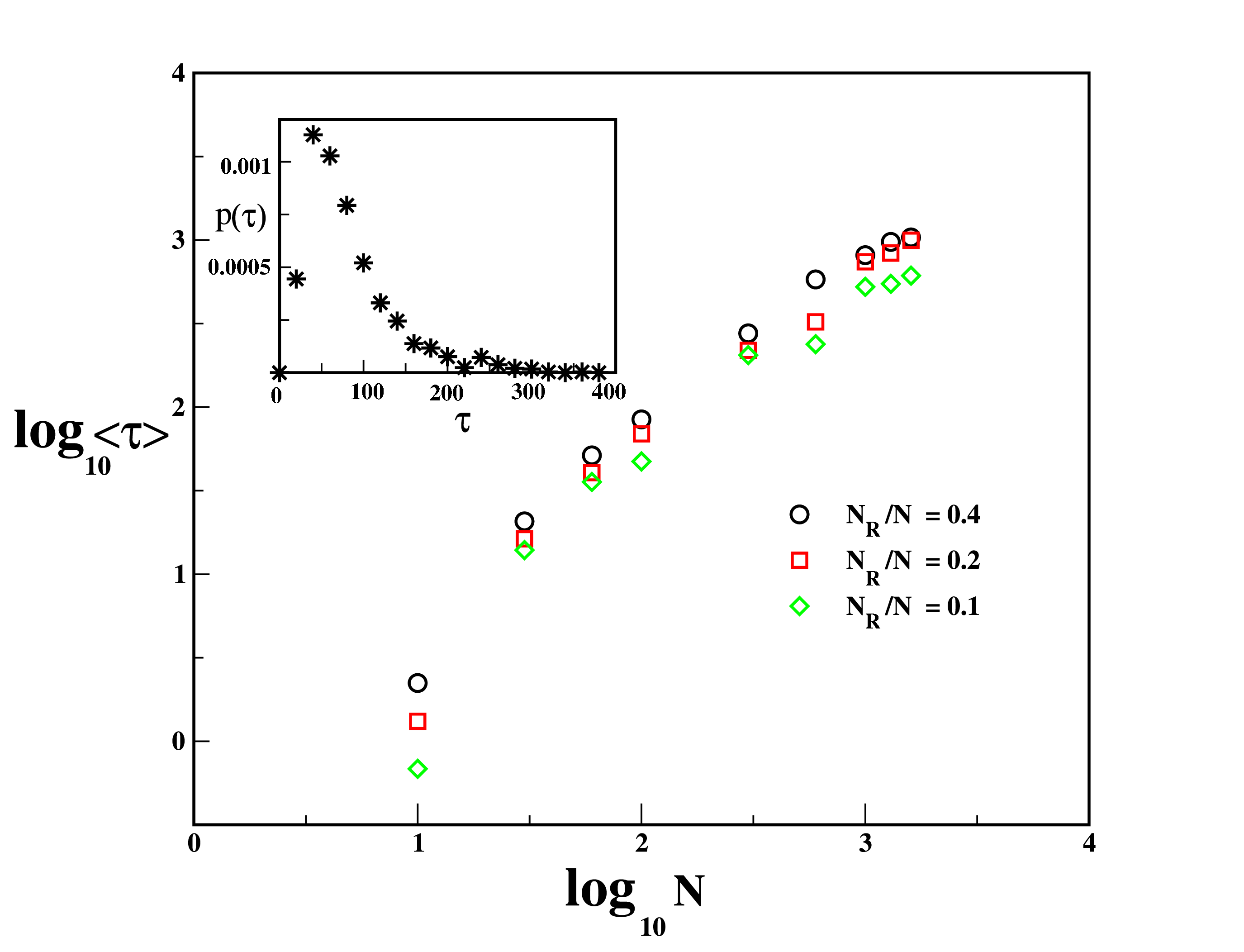}
\caption{Scaling of the mean time for chirality discrimination with total number of reactants for different initial data, showing roughly $\langle \tau \rangle \propto N$. Inset shows the distribution of $\tau$
has an exponential tail for a fixed initial data, $(N_A=100, N_R=10)$, averaged over 500 realisations to obtain smoothened data.}
\label{taumean}
\end{center}
\end{figure}

\begin{figure}
\begin{center}
\includegraphics[width=4in]{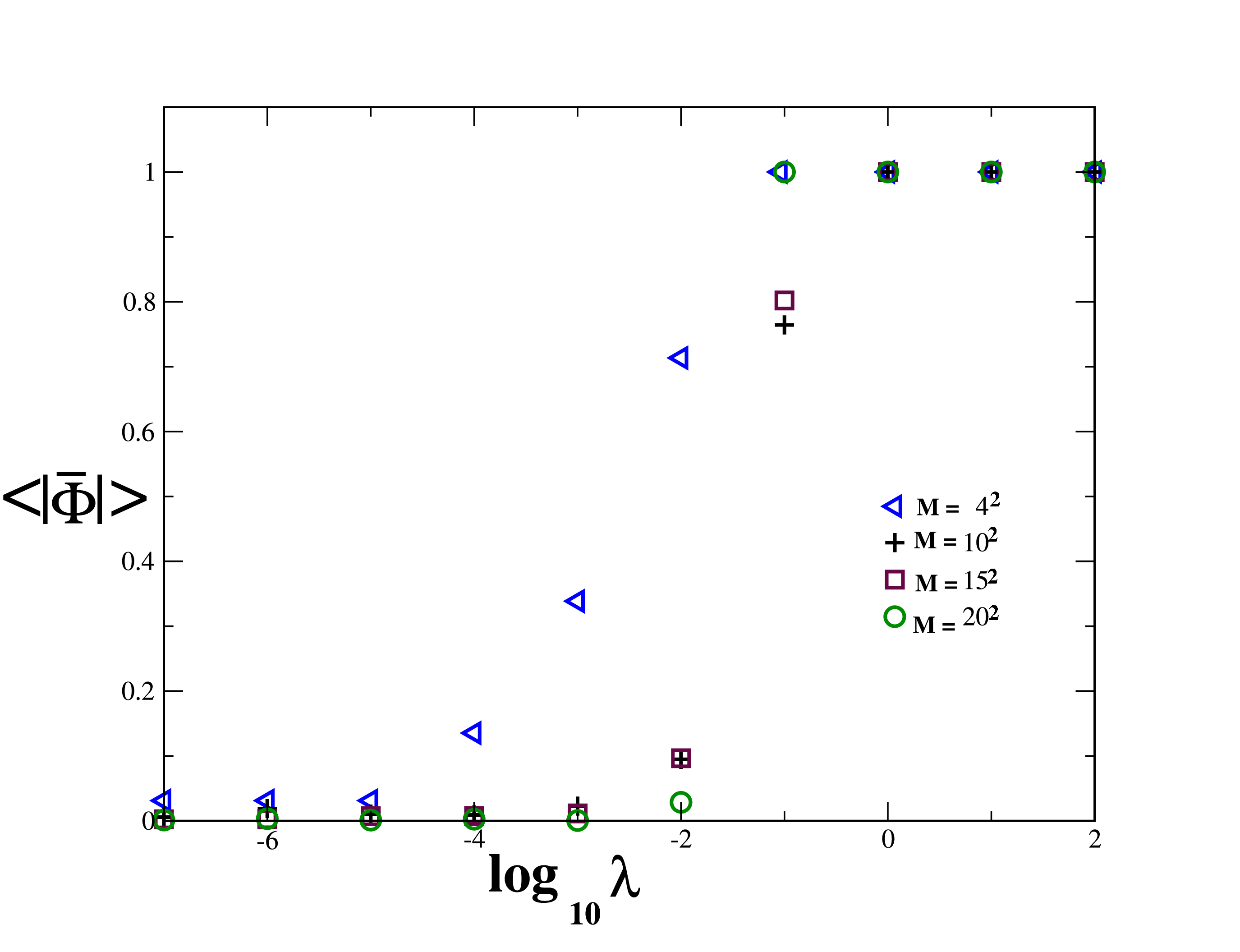}
\caption{Saturated value of the magnitude of the spatially averaged enantiomeric excess as a function of
$\lambda$ shows a jump at $\lambda^*$ which gets sharper with increasing $M$.}
\label{orderparam}
\end{center}
\end{figure}

\begin{figure}
\begin{center}
\includegraphics[width=4in]{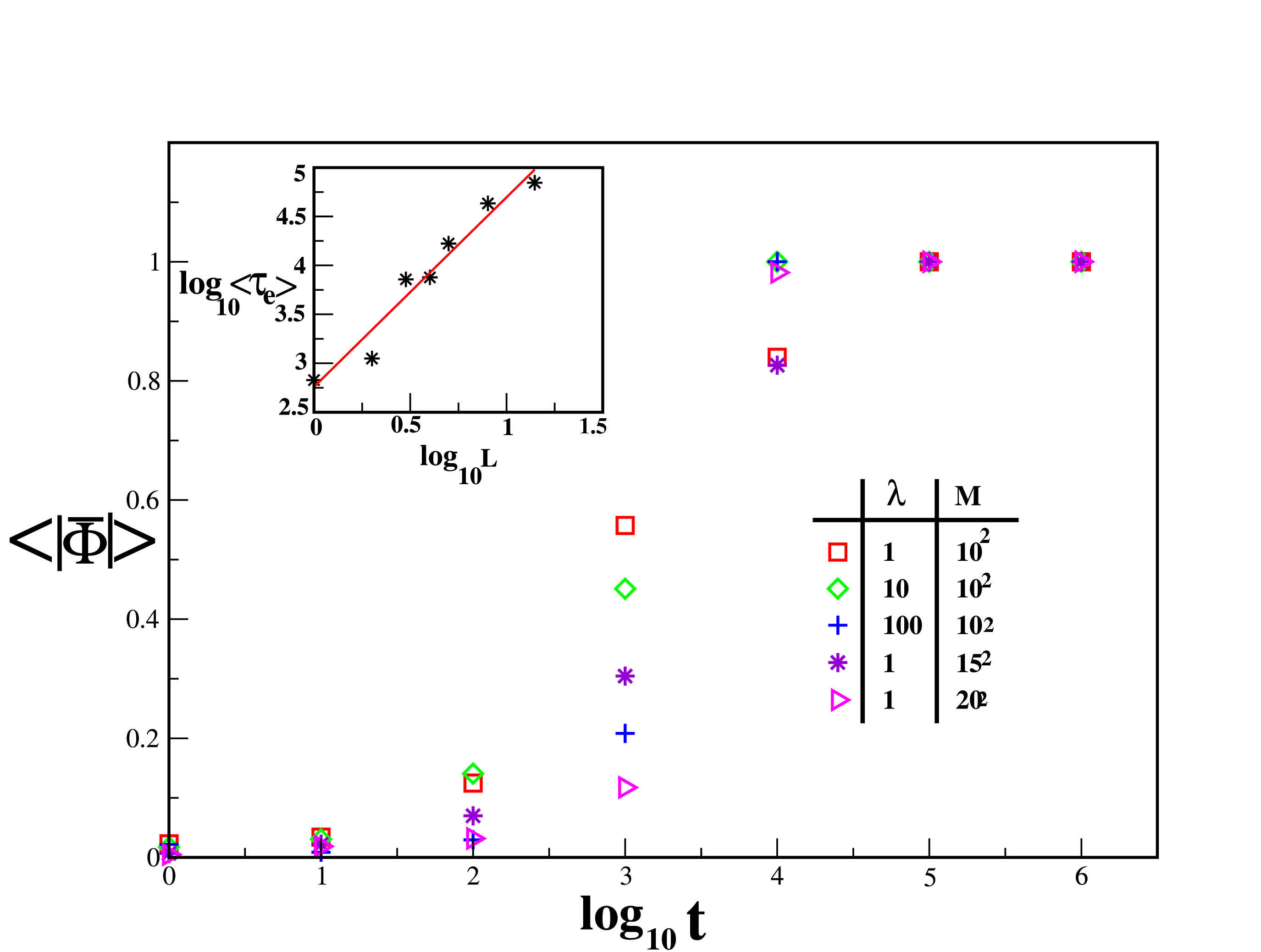}
\caption{Time evolution of $\langle \vert {\bar \Phi}\vert \rangle$ showing typical nucleation and growth, for various system sizes $M$ and diffusion rate $\lambda$. Note that the nucleation time is independent of $M$ and decreases with increasing $\lambda$. Inset shows the scaling of the `evaporation' time of a single droplet of size $L$, $\langle \tau_e \rangle \sim L^{z}$, a best fit gives $z \approx 1.93 \pm 0.2$.}
\label{orderparam}
\end{center}
\end{figure}


\end{widetext}


\begin{thebibliography}{}

\bibitem{note1} Contrast this with noise induced symmetry breaking transitions in nonequilibrium systems with multiplicative noise \cite{noise}, where the symmetry breaking can occur even in the absence of noise provided one spans the entire range of values for the control parameter.

\bibitem{noise} C.Van den Broek, J.M.R. Parrondo and R. Toral, Phys. Rev. Lett. {\bf 73}, 3395  (1994);
W. Horsthemke and R. Lefever, {\it Noise-Induced Transitions}, (Springer-
Verlag, Berlin, 1984).


\bibitem{frank} F. C. Frank, Biochim. Biophys. Acta. {\bf 11}, 459 (1953).

\bibitem{soai} K. Soai et al., 
Nature {\bf 378}, 767 (1995);
T. Shibata et al., Chem. Commun., 1235 (1996);
T. Shibata et al., J. Am. Chem. Soc. {\bf 118}, 471 (1996).

\bibitem{blackmond} D. G. Blackmond, Proc. Natl. Acad.  Sc.
{\bf  101}, 5732 (2004).

\bibitem{noteamp} This situation is different from the excess enantiomeric amplification observed in \cite{frank} and Ya. B. Zeldovich and A.S. Mikhailov, Sov. Phys. Usp. {\bf 30}, 977 (1988).

\bibitem{bykov}Note that this differs from the role of multiplicative noise in, V.I. Bykov et al., Reaction and Catalyisis Letters {\bf 15}, 55 (1980).

\bibitem{gillespie} D. T. Gillespie, J. Phys. Chem., {\bf 81}, 2340 (1977).

\bibitem{marion} G. Marion, X. Mao, E. Renshaw and J. Liu, Phys. Rev. E {\bf 66}, 051915 (2002).

\bibitem{future} V.S. Gayathri and M. Rao, in preparation.

\bibitem{parrondo} For a review of Parrondo's games see, G.P. Harmer and D. Abbott, (2002)
Fluctuation and Noise Lett. {\bf 2}, R71 (2002); J. Buceta, K. Lindenberg and J.M.R. Parrondo, 
Phys. Rev. E {\bf 66}, 036216 (2002).

\bibitem{kondepudi} D. K. Kondepudi, R. Kaufmann and N. Singh, Science {\bf 250}, 975-976  (1990); K. Asakura et al., J. Phys. Chem. A {\bf 104}, 2689 (2000).

\bibitem{metcalfe} G. Metcalfe and J. M. Ottino, Phys. Rev. Lett. {\bf 72},  2875 (1994).

\bibitem{eigen} M. Eigen, J. McCaskill and P. Schuster, J. Phys. Chem. {\bf 92}, 6881 (1988).

\bibitem{kimura} M. Kimura in {\it The neutral theory of molecular evolution}, (Cambridge University Press, 1983).

\end{thebibliography}
\end{document}